\documentclass[twocolumn,showpacs,preprintnumbers,amsmath,amssymb]{revtex4}
\usepackage{graphicx}% Include figure files
\usepackage{dcolumn}% Align table columns on decimal point
\usepackage{bm}% bold math

\begin{document}

\title{Quantum Hall line junction with impurities as a multi-slit Luttinger liquid interferometer}

\author{
I. Yang$^{*}$, W. Kang$^{*}$, L.N. Pfeiffer$^{\S}$, K.W. Baldwin$^{\S}$,\\
 K.W. West$^{\S}$, Eun-Ah Kim$^{+}$, and E. Fradkin$^{+}$}

\affiliation{$*$James Franck Institute and Department of Physics, University of Chicago, Chicago, IL 60637\\
${\S}$Bell Laboratories, Lucent Technologies, 600 Mountain Avenue, Murray Hill, NJ 07974\\
$^{+}$Department of Physics, University of Illinois at Urbana-Champaign, Urbana, IL 61801}

\date{\today}

\begin{abstract}
We report on quantum interference between a pair of counterpropagating quantum Hall edge states that are separated by a high quality tunnel barrier.  Observed Aharonov-Bohm oscillations are analyzed in terms of resonant tunneling between coupled Luttinger liquids that creates bound electronic states between pairs of tunnel centers that act like interference slits. We place a lower bound in the range of 20-40 $\mu$m for the phase coherence length and directly confirm the extended phase coherence of quantum Hall edge states.
\end{abstract}


\maketitle

Two-dimensional electron systems  under strong
magnetic fields condense into incompressible electron liquid states characterized by a
rational value of Hall conductance $\sigma_{xy} = e^{2}/h$.\cite{Prange90, DasSarma97}
The physical excitations of the boundary of these incompressible fluid states are gapless and propagate ballistically parallel to the confining edge potential along a direction determined by the magnetic field. As the only active conduction channel of quantum Hall droplets,
 the edge states of a single isolated quantum Hall fluid are able to skirt local potential defects and to transport electrical current without backscattering.\cite{Komiyama89,Main94,Johnson95,Machida96,Liu98,Acremann99,Roddaro03} The edge states of quantum Hall systems thus constitute a nearly ideal one-dimensional electronic system supporting coherent quantum transport of electrons over large distances.\cite{Wen90,KaneFisher,Chang03} Due to the chiral nature of the edge excitations, the coherence length of the edge excitations is expected to be extremely long and only limited by inter-edge backscattering processes. 

In the fractional quantum Hall regime the edge states of a quantum Hall fluid depart drastically from that of a simple Fermi liquid and behave instead as chiral Luttinger liquids.\cite{Wen90,KaneFisher} 
However,  the edge states of two quantum Hall fluids in the integer regime can also exhibit non-Fermi liquid behavior  if brought sufficiently close to each other. Due to the effects of inter-edge correlations,  two strongly coupled, counterpropagating edge states behave as a single non-chiral Luttinger liquid whose Luttinger parameter $K$ and propagation velocity $v$ are continuously tuned by the magnetic field, leading to a drastic modification of the expected transport properties of this one-dimensional ``wire".\cite{Kim03} 
The inter-wire correlation in coupled Luttinger liquids has led to predictions of striking quantum effects.\cite{Kim03,Chamon97a,Mitra01,Kim03a,Geller97} 

In this paper, we report on quantum interference effects between two coupled chiral Luttinger liquids formed across a quantum Hall line junction of two-dimensional electron systems separated by a high quality tunnel barrier.  In the limit of zero temperature, the system enters a coherent tunneling regime where the condition for quantum interference can be realized and the conductance across the line junction exhibits the characteristic set of oscillations associated with the Aharonov-Bohm (AB)  interference. We interpret these AB oscillations as the signature of a resonant state created by two tunneling centers that strongly couple the counter-propagating edge states like a slit for quantum interferometer. Presumably these centers are created by a few widely separated small defects or impurities whose role is to strongly couple the two counterpropagating edge states through enhanced tunneling at these sites. From the period of the oscillations, we determine the size of the AB trajectories and establish a lower limit of 20-40 $ \mu$m as the minimum phase coherence for edge electrons. Our results confirm the expectation that the quantum Hall edge states possess an enormously large phase coherence length.

\begin{figure}
\begin{center}
\includegraphics[width=\linewidth]{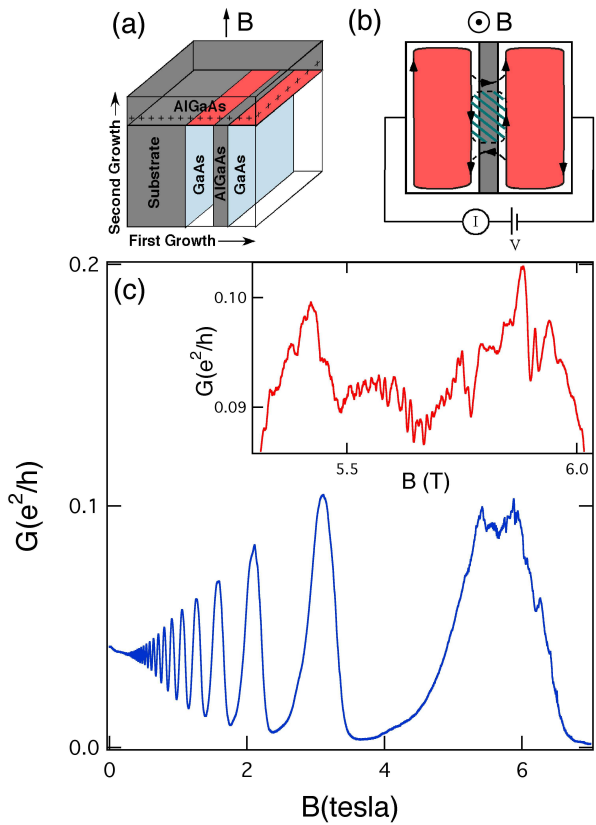}
\end{center}
\caption{\label{Fig:ab2}
(a) Schematic of the line junction tunneling structure based on the cleaved edge overgrowth.   (b) Edge state and AB trajectories in the quantum Hall line junction under quantizing magnetic field.  (c) Magnetic field sweep at 25 mK. 
 Inset: An expanded view of the conductance fluctuations of the final conductance peak at 5.6 Tesla.} 
\end{figure}

The line junctions are grown by cleaved edge overgrowth
using molecular beam epitaxy (MBE) on the (110) face of GaAs/AlGaAs multilayer structure.\cite{Pfeiffer90,Kang00} Figure~\ref{Fig:ab2}a 
illustrates the layout of the line junction device.  The initial growth along the (100) direction consists of undoped 13 $\rm{\mu m}$ GaAs followed by a 8.8 nm-thick alloy of Al$_{0.3}$Ga$_{0.7}$As and completed by 14 $\rm{\mu m}$ layer of undoped GaAs. The multilayer structure is then cleaved along the (110) plane and a modulation-doping is performed over the exposed edge, forming two side-by-side sheets of two-dimensional electrons separated from each other by the  Al$_{0.3}$Ga$_{0.7}$As barrier. Independent contacts to individual two-dimensional electrons were made using evaporated AuGeNi. 
Incommensurate conductance fluctuations were detected in total of three samples. For the consistency the data shown throughout this paper are from one sample whose density of the two-dimensional electron was $n \approx 2\times 10^{11} \rm{cm^{-2}}$ with a mobility of $\sim 1\times 10^{5} \rm{cm^{2}\,V^{-1}\,sec^{-1}}$. 
Figure~\ref{Fig:ab2}b illustrates the expected edge state trajectories under magnetic field and the measurement geometry. 

Figure~\ref{Fig:ab2}c shows the magnetic field dependence of the differential conductance $G$, at zero bias under 25 mK of temperature. The zero bias tunneling conductance (ZBC) peaks successively grows in magnitude with magnetic field until reaching the final conductance peak centered around 5.6 tesla.
Above 7 tesla, the conductance becomes vanishingly small as the condition for transverse momentum conservation suppresses tunneling at zero bias in the fractional quantum Hall regime.\cite{Kang00,Mitra01,Ho94,Takagaki00,Nonoyama02} The last conductance peak exhibits conductance fluctuations arising from Aharanov-Bohm oscillations. 
The inset of Fig.~\ref{Fig:ab2}c shows  an expanded view of the oscillations in the conductance around the maximum. We find a set of small period oscillations superimposed on top of irregular features at larger magnetic field scales. The larger period structures are highly irregular and generally distort  the shape of the peak. 
Applying a finite bias to the one side of the junction with respect to the other sharply depresses these oscillations. We interpret this behavior as a consequence of the electron heating inducing suppression of AB oscillations under finite bias conditions. 

Two different theoretical scenarios have been proposed to describe the physics of the line junction in the clean limit. Within the level-mixing picture of tunneling across the line junction,\cite{Mitra01,Ho94,Takagaki00,Kollar02,Nonoyama02} the conductance peaks occur whenever the energy levels of the two edges coincide with the Fermi level at zero bias as a function of magnetic field. It was also proposed that the ZBC peak arises from the formation of a correlated electronic state with spontaneous inter-edge coherence at zero momentum transfer.\cite{Mitra01}
In an alternate picture, the ZBC peak is due to the effects of point-contact tunneling in the Coulomb-coupled edge states.\cite{Kim03}
In this framework the successive ZBC peaks are due to quantum phase transitions tuned by the magnetic field, caused by opening and closing of tunneling channels  between the coupled edge states as the magnetic field is varied.  In contrast to the Landau level mixing mechanism, the point-contact mechanism provides a natural mechanism for the  AB effect, provided that there are multiple tunneling centers embedded within the barrier.\cite{Kim03a} (The case for the dirty limit was addressed by Ref.~\onlinecite{Kane97}). The AB oscillations that we report here are consistent with this interpretation.

\begin{figure}
\includegraphics[width=\linewidth]{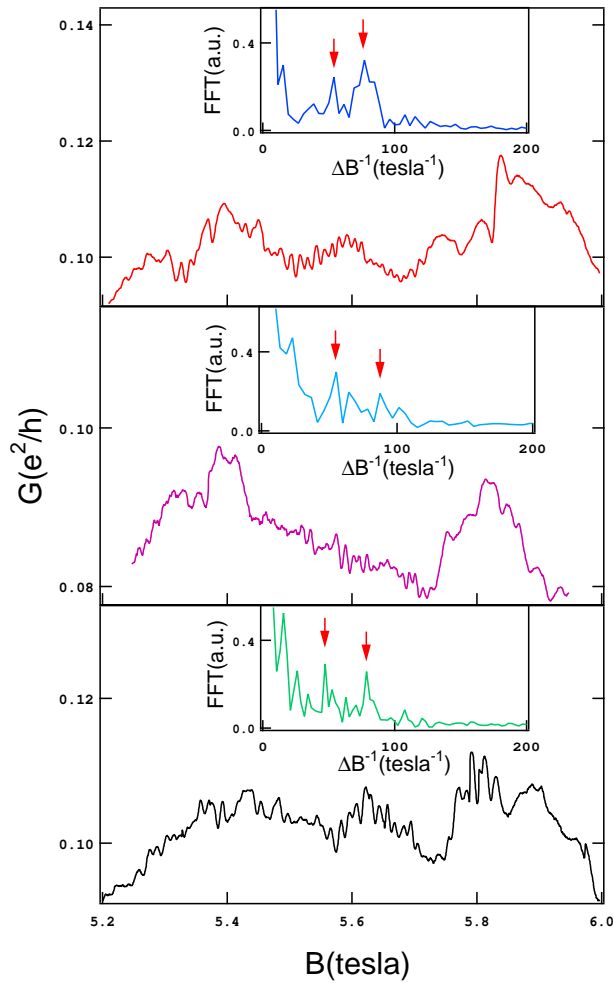}
\caption{\label{Fig:ab3} Quasiperiodic conductance oscillations obtained from different thermal cyclings. 
Insets: Fast Fourier transforms of the conductance data demonstrate that at least 2 large periods are visible in the conductance.} 
\end{figure}

Figure~\ref{Fig:ab3} illustrates the conductance fluctuations detected near the final conductance peak centered around 5.6 tesla. In all cases conductance exhibits a reproducible set of small oscillations superimposed on slowly varying oscillations. The small period oscillations can change depending on history and thermal cyclings. Once cooled to low temperatures, the conductance fluctuations become robust and reproducible. Visually, these oscillation are quasi-periodic and demonstrate beating from presence of multiple frequencies. The insets of Fig.~\ref{Fig:ab3} show  the result of fast fourier transform (FFT) analysis of the conductance traces, yielding at least 2 primary frequencies in addition to other, small amplitude frequencies. Inverse FFT shows that the conductance oscillations are predominantly determined by the 2 primary frequencies with negligible contribution coming from the small amplitude frequencies.   In case of the top conductance trace in Fig.~\ref{Fig:ab3}, the slowly varying oscillation of $\sim$0.2 tesla is complemented by the quasi-periodic oscillations derived from at least two distinct frequencies of 53.8 and 77.0 tesla$^{-1}$, corresponding to the periods of $\Delta B$ = 13.0 and 18.6 millitesla respectively. 

Appearance of conductance oscillations in the final conductance peak near $\nu = 1$ suggests that the Aharnov-Bohm effect occur in the strong tunneling regime discussed in Ref.~\onlinecite{Kim03a}. As electrons propagate parallel to the barrier, the tunneling hotspots define a set of Feynman paths that encircle an area defined by the distance between the tunneling hotspots, $a$, and the width of the barrier, $d$, as illustrated in Fig.~\ref{Fig:ab2}b. In this regime, a system of two coupled chiral edges with two tunnelling centers behaves as two semi-infinite Luttinger ``leads" coupled through an elongated island, qualitatively defined by a set of closed Feynman paths. 
In this situation, first described by Kane and Fisher in the context of quantum wires,\cite{kane-fisher} conduction along the barrier proceeds through resonant hopping processes through the island. In the particular case of this quantum Hall system, the resonance condition is tuned by the external magnetic field due to the chiral nature of the edges. The resonance condition is met when the flux enclosed in the island is a half-integer multiple of the flux quantum $\phi_0=h/e$. Near a resonance, the tunneling conductance across the barrier is strongly suppressed leading to the observed sharp AB-like oscillations.   

\begin{figure}
\includegraphics[width=\linewidth]{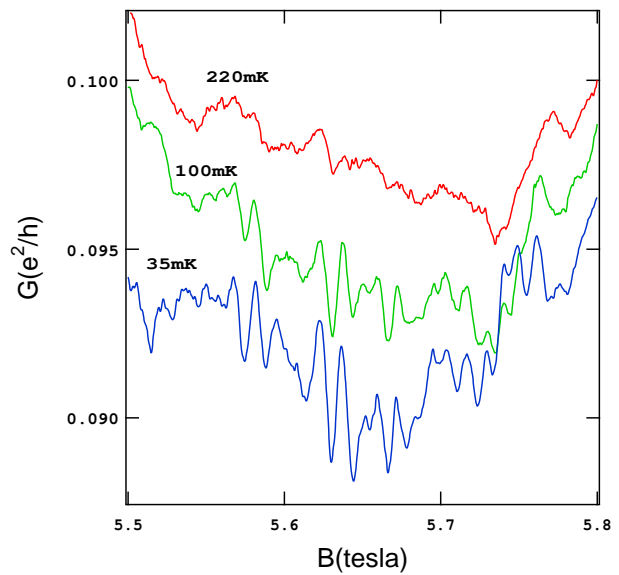}
\caption{\label{Fig:ab4} Temperature evolution of the conductance 
fluctuations. Small period oscillations disappear above 200 mK. } 
\end{figure}

It has been proposed that conductance oscillations due to AB effect of anti-dot structure in the quantum Hall regime\cite{Goldman95,Kataoka99} are mediated by Coulomb blockade of electrons around the anti-dot.\cite{Kataoka99} It must be emphasized that the AB effect in the line junction does not involve Coulomb blockade. Formation of a contiguous interference trajectory requires tunneling through two locations in the barrier while maintaining phase memory at the same time. Depending on the distance separating these tunneling centers, electrons can coherently tunnel back and forth if the phase-coherence time of edge electrons is greater than the thermal decoherence time. At higher temperatures, thermal broadening is expected to suppress the coherence of the AB oscillations. Figure~\ref{Fig:ab4} shows the effect of increasing temperature on these oscillations. The small period oscillations are considerably weakened by 100 mK and largely disappear above 200 mK. Only the weak remnants of larger period oscillation are visible at higher  temperatures. Such a rapid suppression of the AB oscillations with increasing temperature shows that the observed oscillations are the result of quantum mechanical phase-coherence.

A realistic barrier possessing more than two point contacts will produce AB oscillations with a complex interference patterns due to many possible interference pathways. These ``point contacts" may be the sites that contain an impurity or a defect that enhances tunneling at the particular location. Presence of two distinct oscillation frequencies, as obtained from the FFT analysis, suggests that there are two primary interference pathways likely established by at least three preferential tunneling spots or alternatively two pairs of resonant states along the length of the junction. The oscillation period $\Delta B$ for an enclosed area $A$ is given by $\Delta B = \frac{h}{e}\frac{1}{A}$. The interference pathways, as defined by the distance between the interference slits and the width of the tunnel barrier, provide a measure of the phase coherence length of the tunneling electrons. Table~\ref{Table1} summarizes the periods of oscillations and the corresponding distances traveled by the tunneling electrons parallel to the barrier assuming that the electrons travel immediately next to the barrier. The periods of oscillations $\Delta B$ = 11-13 and 18-21 millitesla determined from the FFT analysis of the dataset yield corresponding lengths of 36 - 41 and 22 - 26 $\rm{\mu m}$, respectively. Since the actual phase coherent length is larger than these distances, these lengths provide a measure of the lower bound for the phase coherence length in  the prescribed geometry.

 \begin{table}
\caption{Periods of principle oscillations, $\Delta$B$_{1}$ and $\Delta$B$_{2}$ from the fast Fourier transform of the conductance oscillations from Fig.~\ref{Fig:ab4}, and the corresponding distances, $a_{1}$ and $a_{2}$, between the interference sites along the junction.}
\begin{tabular}{ccccc}
\hline\hline
Data	 set & $\Delta$B$_{1}$(mT) & $\Delta$B$_{2}$(mT) & $a_{1}$($\mu$m) & $a_{2}$($\mu$m) \\
\hline
a & 13.0 & 18.6 & 36.2  & 25.3   \\
 b  & 11.4 & 18.1  & 41.1  & 26.0   \\             
 c & 12.7 & 21.2 & 37.0 & 22.2   \\           
\hline\hline
\end{tabular}
\label{Table1}
\end{table}

The lower bound for the phase coherent length may be smaller since the distance between the two counterpropagating paths that produce the AB interference may be a little wider than the barrier width, proportionally reducing the distances estimated in Table~\ref{Table1}.  Since the guiding center for the zero-bias conductance states lie at the center of the barrier, it is unlikely that the distances between counterpropagating trajectories of the tunneling electrons will approach two or three times the magnetic length. If the separation between opposite legs of the AB trajectory is doubled to account for the uncertainty in the shape of the trajectories, then the lower bound of the phase coherence length is reduced to about $\sim$20 $\mu$m, which still remains a substantial phase coherence length in a solid state environment. 

Our estimate of the phase coherence length compares favorably with the measurement of the zero-field electronic phase coherence length of 4-10 $\mu$m for GaAs/AlGaAs heterostructures determined from weak-localization analysis of transport in lithographically defined one-dimensional channels.\cite{Kurdak92,Katine98} The phase coherence length from weak localization analysis was performed for samples with mobilities comparable to our sample.  While the bulk 2D electron system in our sample possesses relatively low mobility, the tunnel barrier possesses very little disorder and this should sustain ballistic transport of electrons parallel to the barrier. The fact that the edge state in our device can exhibit such a large phase coherence length in spite of the moderate bulk mobility attests to the remarkable transport properties of quantum Hall edge states. On a related note, our measure of the lower bound on the phase coherence length is about 100 times smaller than the length of 5.4 mm determined from an earlier experiment on the narrowing of the transition between two phase-separated regions in a Hall bar.\cite{Machida98} This claim was never verified independently. Our experiment differs from Ref.~\onlinecite{Machida98} in that our determination of the lower bound of the phase coherence length is based on an explicit detection of quantum interference. 
Our estimate of the coherence length is comparable to the circumference of the electronic interferometer by Ji. {\em et al.} where AB effect is observed from the change of magnetic flux in an area of $\sim$~45 $\mu$m$^2$ enclosed by two separated electronic paths.\cite{Ji03} 
In line junctions with  higher bulk mobility, it should be possible to establish an even larger bound for the phase coherence length.

In summary, we have observed AB effect arising from quantum interference between two counterpropagating edge states across a quantum Hall line junction. The observed AB  oscillation is understood in terms of resonant tunneling between coupled chiral Luttinger liquids that creates bound electronic states across the line junction. The formation of the bound states is  mediated by impurities in the barrier that act like interference slits. From the periods of conductance fluctuations, our conservative estimate places a range of 20-40 $\rm{\mu m}$ as the phase coherence length for quantum Hall edge states.  As the actual phase coherent length is longer than the distances between the tunnel sites, it is probable that the edge electrons in the line junction are able to maintain phase coherent motion in excess of 20 $\rm{\mu m}$ as it moves along the junction.
  
The work at the University of Chicago is supported by NSF DMR-0203679 and NSF MRSEC Program under DMR-0213745. The work at the University of Illinois (EAK and EF) was supported in part by the National Science Foundation through the grant DMR01-32990.

\end{document}